\documentclass[a4paper,12pt]{article}

\usepackage{graphicx}%

\begin{document}

\section{Copyright}

This is a preprint version of an article submitted to the Journal of Mathematical Chemistry, available at \\
www.springerlink.com

\title{
Towards an exact orbital-free single-particle kinetic energy density for the inhomogeneous electron liquid in the Be atom.
}
\author{by 
A. Krishtal$^{1}$, N.H. March$^{2,3,4}$, and C. Van Alsenoy$^{1}$}
\date{}
\maketitle
\begin{center}
{
$^{1}$Chemistry Department, University of Antwerp, Antwerp, Belgium.\\
$^{2}$Physics Department, University of Antwerp, Antwerp, Belgium.\\
$^{3}$Oxford University, Oxford, England.\\
$^{4}$Abdus Salam International Centre for Theoretical Physics, Trieste, Italy
}
\end{center}

\begin{abstract}
\noindent 
Holas and March (Phys.\ Rev.\ {\bf A51}, 2040 (1995)) wrote the gradient
of the one-body potential V(r) in terms of low-order derivatives of the
idempotent Dirac density matrix built from a single Slater determinant
of Kohn-Sham orbitals. Here, this is first combined with the study of
Dawson and March (J.\ Chem.\ Phys.\ {\bf 81}, 5850 (1984)) to express the
single-particle kinetic energy density of the Be atom ground-state in terms of
both the electron density $n(r)$ and potential $V(r)$. While this is the more compact formulation, we then, by removing $V(r)$, demonstrate that the ratio $t(r)/n(r)$ depends, though non-locally, only on the single variable $n'(r)/n(r)$, no high-order gradients entering for the spherical Be atom.
\end{abstract}

\noindent Keywords : electron density $n$, gradient quotient $n'/n$, orbital-free kinetic energy

\newpage

\noindent 
In recent work, we have been focussing on approximations to the
exchange energy density in DFT \cite{ref1}. In this area, the long term
aim is to express the single-particle kinetic energy $T_S$ and the
exchange energy $E_x$ (plus later correlation energy) as functionals of
the ground-state electron density \cite{ref2}. A recent review on such an orbital-free DFT has been presented by one of us in this Journal.\cite{ref3_new}

\noindent
Here, we focus primarily on the kinetic energy $T_S$ which we shall
write in the form
\begin{equation}
\label{eq:1}
T_S = \int t(r) d\vec{r}
\end{equation}
where $t(r)$ is evidently the single-particle kinetic energy
density. Throughout this Letter, we shall use the
positive-definite form of $t(r)$, corresponding to $(\nabla
\Psi)^2$ in terms of the one-electron wavefunctions.

\noindent
The general problem of finding $T_S[n]$ remains formidable. Therefore our
focus here is on the example of the spherically symmetric Be atom ground
state, with the configuration (1s)$^2$(2s)$^2$. Then, we know from the
early study of Dawson and March \cite{ref3} that the kinetic energy
density $t(r)$ entering in eqn.\ (\ref{eq:1}) can be usefully separated
into two parts :
\begin{equation}
\label{eq:2}
t(r) = t_W(r) + \frac{1}{2}\ n(r)\ \theta^{'2}(r)
\end{equation}
where $t_W(r)$ is the von Weizs\"acker \cite{ref4} kinetic energy density
given by
\begin{equation}
\label{eq:3}
  t_W(r) = \frac{1}{8}\ \frac{(\nabla n)^2}{n} 
\end{equation}
So from eqns.\ (\ref{eq:2}) and (\ref{eq:3}) it is already clear that the
sought-after functional form of $T_S[n]$ must involve both the density
$n$ and the gradient-dependent ratio $\nabla n/n$.

\noindent
Returning to \cite{ref3}, Dawson and March wrote the 1s and 2s
Slater-Kohn-Sham \cite{ref5,ref6} orbitals in the forms
\begin{equation}
\label{eq:4}
2^{\frac{1}{2}}\Psi_{1s}(r) = n^{\frac{1}{2}} cos \theta(r)
\end{equation}
and
\begin{equation}
\label{eq:5}
2^{\frac{1}{2}}\Psi_{2s}(r) = n^{\frac{1}{2}} sin \theta(r)
\end{equation}
for normalised $\Psi$'s. But, returning to eqn.\ (\ref{eq:2}), the most
interesting result is a non-linear pendulum-like differential equation
for the phase $\theta(r)$: namely
\begin{equation}
\label{eq:6}
\frac{}{}\nabla^2 \theta(r) + \frac{\nabla n}{n}.\nabla \theta(r) + \lambda
sin 2\theta(r) = 0
\end{equation}
It is immediately clear from eqn.\ (\ref{eq:6}) that the only input
required to determine $\theta(r)$ in the Be atom is the gradient ratio 
$\nabla n/n$ already mentioned in connection with the von Weizs\"acker
component $t_W(r)$ in eqn.\ (\ref{eq:3}). Then the eigenvalue $\lambda$, having dimensions of energy,
and the phase $\theta(r)$ must be determined by integrating (usually
numerically!) the non-linear (ordinary, for spherical atoms like Be or
4-electron atomic ions) differential eqn.\ (\ref{eq:6}).

\noindent
Such integration presumably will mean that while eqn.\ (\ref{eq:6}) shows
already that $\theta$ is a functional of $\nabla n/n$, i.e.
\begin{equation}
\label{eq:7}
\theta = \theta[\nabla n/n]
\end{equation}
the dependence of $\theta$ on $r$ will not be determined by the `local'
quantity $\nabla n(r)/n(r)$ at the point $r$ at which $\theta$ is to be
evaluated. Thus, eqn.\ (\ref{eq:7}) is to be interpreted as a `non-local'
functional of the variable $\nabla n/n$.

\noindent
We turn at this point to introduce the one-body potential V(r).
This enters, most basically in DFT \cite{ref2}, through the chemical
potential ($\mu$) equation
\begin{equation}
\label{eq:8}
\mu = \frac{\delta T_S}{\delta n(r)} + V(r).
\end{equation}
Of course, in current applications of DFT, the functional derivative
$\delta T_S/\delta n(r)$ is bypassed by utilizing the Slater-Kohn-Sham (SKS)
orbitals generated by the potential $V(r)$\cite{ref2}. Unfortunately, this potential
has an additive component from exchange and correlation, $V_{xc}(r)$,
which is currently not known.

\noindent
For Be, we shall also bypass eqn.\ (\ref{eq:8}) because of the difficulty
associated with forming $\delta T_S/\delta n(r)$ after inserting the
formally exact eqn.\ (\ref{eq:1}). What we appeal to instead is the
density matrix formulation of DFT given by Holas and March\cite{ref7}
and used very recently by March and Nagy\cite{ref8} in conjunction with
the concept of the Pauli potential\cite{ref9}. From eqn.\ (\ref{eq:1}) of \cite{ref7}
we can write in spherical symmetry the force equation
corresponding to the one-body potential V(r) :
\begin{equation}
\label{eq:9}
-\frac{\partial V(r)}{\partial r} = -\frac{1}{4}
\frac{(\frac{\partial}{\partial r})[\nabla^2 n(r)]}{n(r)} + \frac{z_s(r)}{n(r)}
\end{equation}
The quantity $z_s(r)$ entering eqn.\ (\ref{eq:9}) is the single-particle
$(s)$ limit of the many-electron vector field $\vec{z}(\vec{r})$ defined
in \cite{ref7} and written explicitly in eqns.\ (2) and (3) of
\cite{ref8}.

\noindent
Utilizing the idea of the Pauli potential, in \cite{ref8} it is shown,
but now specifically for the Be atom, that
\begin{equation}
\label{eq:10}
z_s(r) = \frac{4t(r)}{r}\ +\ 2\frac{\partial t(r)}{\partial r}
\end{equation}
where $t(r)$ is given in eqn.\ (\ref{eq:2}) above. Thus, at the expense
of working in eqn.\ (\ref{eq:9}) with the force, rather than $V(r)$ in the
chemical potential result (\ref{eq:8}), we have bypassed the need for
functional differentiation of $T_s[n]$ : of course the result
(\ref{eq:10}) being specific for the two-level Be atom.\cite{ref10}

\noindent
We must note here that in the very early work on the virial
theorem in one dimension, \cite{ref10} it was shown that a differential virial form
existed, namely
\begin{equation}
\label{eq:11}
\frac{\partial t}{\partial x} = -\frac{1}{2} n(x) \frac{\partial
V(x)}{\partial x} + \frac{1}{8}n^{'''}(x)
\end{equation}
It was clear already from the one-dimensional form (\ref{eq:11}) that
$t(x)$ was very compactly expressed in terms of both $n(x)$ and its
derivatives, plus the force $-\partial V(x)/\partial x$. Eqn.\
(\ref{eq:10}), when inserted into eqn.\ (\ref{eq:9}), leads to a related
conclusion for the Be atom.

\noindent However, to achieve the prime object of this Brief Report, we want now,
despite some mathematical complications, to avoid the force term
$-\partial V(r)/\partial r$ on the LHS of eqn.\ (\ref{eq:9}). To do so,
let us return to eqn.\ \ref{eq:2} and define a kinetic energy difference $\Delta(r)$ by
\begin{equation}
 \label{eq:12}
\frac{2[t(r)-t_W(r)]}{n(r)}=\Delta(r)=\theta^{'2}(r)
\end{equation}
Taking the square root of eqn.\ (\ref{eq:12}) and noting that $\theta^{'}(r)$ for Be is negative while $\Delta(r)$ is greater than zero, we obtain by integration that
\begin{equation}
 \label{eq:13}
\theta(r)=-\int^{r}\Delta^{\frac{1}{2}}(s)ds
\end{equation}
This is the point at which to return to the non-linear pendulum-like eqn.\ (\ref{eq:6}). Written for the spherically symmetric Be atom, we then find
\begin{equation}
 \label{eq:14}
\theta^{''}(r)+\{ q+\frac{2}{r} \} \theta^{'}(r)+\lambda\sin{2\theta(r)}=0
\end{equation}
where $q$ is the gradient quotient $n^{'}(r)/n(r)$.

\noindent Multiplying eqn.\ (\ref{eq:14}) throughout by $\theta^{'}$ then yields
\begin{equation}
 \label{eq:15}
\theta^{'}(r)\theta^{''}(r)+\{q+\frac{2}{r}\}\theta^{'2}{r}+\lambda\theta^{'}(r)\sin{2\theta(r)}\theta^{'}(r)=0
\end{equation}
The first two terms in eqn.\ (\ref{eq:15}) can be written in terms of $\Delta(r)$ defined in eqn.\ (\ref{eq:12}), and its first derivative $\Delta^{'}{r}$, while the third term involves $\theta^{'}{r}=-\Delta^{\frac{1}{2}}(r)$ as well as $\theta(r)$ given in eqn.\ (\ref{eq:13}). The resulting equation then reads
\begin{equation}
 \label{eq:16}
\frac{\Delta^{'}(r)}{2}+\{q+\frac{2}{r}\}\Delta^{'}{r}+\lambda\Delta^{\frac{1}{2}}\sin(-2\int^{r}\Delta^{\frac{1}{2}}(s)ds)
\end{equation}
Evidently, with input of $q=n/n^{'}$ and also $\lambda$, eqn.\ (\ref{eq:16}) can be formally integrated to yield $\Delta(r)$. From eqn.\ (\ref{eq:12}) this will yield the functional form
\begin{equation}
 \label{eq:17}
 \frac{t(r)}{n(r)}=F[q(r);\lambda]
\end{equation}
since from eqn.\ (\ref{eq:3}) in spherical symmetry we have 
\begin{equation}
 \label{eq:18}
 \frac{t_W(r)}{n(r)}=q^2(r).
\end{equation}
Again, as in the formal functional form in eqn.\ (\ref{eq:7}), makes it clear that $t(r)/n(r)$ depends only on the $r$ space variable $q(r)=n^{'}(r)/n(r)$, which of course is a major simplification.

\noindent It is worthy of note that, by a further appropriate differentiation using eqn.\ (\ref{eq:16}), $\lambda$ can be eliminated, but the mathematical detail proliferates and we shall not record it therefore.

\noindent
While this Letter has focussed on the kinetic energy density $t(r)$, we
conclude by noting that the exchange-only energy density
$\varepsilon_{x}(r)$ of DFT is customarily defined, following Dirac,\cite{ref13} in 
terms of an idempotent first-order density matrix
$\gamma(\vec{r},\vec{r}^{'})$. For Be, this has the form in terms of
$\Psi_{1s}(r)$ and$\Psi_{2s}(r)$  written in eqns.\ (\ref{eq:4}) and
(\ref{eq:5}) as
\begin{equation}
\label{eq:19}
\gamma(\vec{r},\vec{r}^{'}) = 2[\Psi_{1s}(r)\Psi_{1s}(r^{'}) + 
\Psi_{2s}(r) \Psi_{2s}(r^{'}])
\end{equation}
But this is well known when using the forms (\ref{eq:4}) and
(\ref{eq:5}) in eqn.\ (\ref{eq:19}) as
\begin{equation}
\label{eq:20}
\gamma(\vec{r},\vec{r}^{'}) = n(r)^{\frac{1}{2}} n(r^{'})^{\frac{1}{2}}
cos[\theta(r) - \theta(r^{'})]
\end{equation}
Evidently since the Dirac\cite{ref11} result for $\varepsilon_x(r)$ in
terms of $\gamma(\vec{r},\vec{r}^{'})$ is
\begin{equation}
\label{eq:a13}
\varepsilon_x(r) = -\frac{e^2}{4} \int
\frac{\gamma(\vec{r},\vec{r}^{'})^2}{|\vec{r}-\vec{r}^{'}|} d\vec{r}^{'}
\end{equation}
all we have done above on $\theta(r)$ for the Be atom is
relevant to the calculation of the exchange energy density
$\varepsilon_x(r)$ in terms of $n(r)$ and the quotient $q(r)=
n^{'}(r)/n(r)$.

\section{Acknowledgments}

NHM completed his contribution to this article at ICTP, Triest, and he thanks Professor V. E. K. Kravtsov for generous hospitality during his stay. NHM wishes to acknowledge recent valuable discussions on the general area embraced in the present study with A.\ Akbari, T.\ G\'{a}l, I.A.\ Howard and A.\ Nagy. NHM also acknowledges partial financial support made possible by Professors D.\ Lamoen and C.V.A. through the University of Antwerp grant BOF-NOI. AK is grateful to the Research Foundation Flanders (FWO) for a postdoctoral position.


\newpage
\begin{thebibliography}{99}

\bibitem{ref1}  N.H. March and C. Van Alsenoy, Phys. Chem. Liquids \textbf{47}, 225 (2009)

\bibitem{ref2} R.G. Parr and W. Yang, Density-Functional Theory of Atoms and Molecules,  Oxford University Press: New York. (1989)

\bibitem{ref3_new} N.H. March, Phys. Chem. Liquids \textbf{48}, 141 (2010).

\bibitem{ref3}  K.A. Dawson, N.H. March, J. Chem. Phys., {\bf 81}, 5850
(1984)

\bibitem{ref4}  C.F. von Weizs\"acker, Z. Phys. \textbf{96}, 431 (1935)

\bibitem{ref5}  J.C. Slater, Phys. Rev. \textbf{81}, 385 (1951).

\bibitem{ref6}  W. Kohn and L. J. Sham, Phys. Rev. \textbf{140}, A1133 (1965).

\bibitem{ref7}  A. Holas, N.H. March, Phys. Rev., {\bf A51}, 2040 (1995)

\bibitem{ref8}  N.H. March, \'{A}. Nagy, Phys. Rev. {\bf A78}, 044501 (2008)

\bibitem{ref9}  N.H. March, J. Mol. Structure: Theochem, in press.

\bibitem{ref10}  N.H. March, W.H. Young, Nucl. Phys., {\bf 12}, 237
(1959)

\bibitem{ref11}  P.A.M. Dirac, Proc. Camb. Phil. Soc., {\bf 26}, 376
(1930)



\end{thebibliography}
\end{document}